\documentclass[prl,twocolumn,showpacs,preprintnumbers,amsmath,amssymb,showkeys]{revtex4}

\usepackage{graphicx}
\usepackage{dcolumn}
\usepackage{bm}
\usepackage{textcomp}
\usepackage{latexsym}
\DeclareMathOperator{\Brill}{{\rm B}_s}

\begin{document}

\preprint{APS/123-QED}

\title{Transverse Magnetoresistance \\ of Zn$_{0.9}$Co$_{0.1}$O:Al Thin Films}

\author{Ricardo Mart\'{i}nez-Valdez}
\altaffiliation[Currently at: ]{Physics Department,\\ University of Puerto Rico, R\'{i}o Piedras}
\author{H\'{e}ctor J.\ Jim\'{e}nez-Gonz\'{a}lez}
\email{hectorj.jimenez@upr.edu}
\affiliation{Physics Department\\
University of Puerto Rico at Mayag\"{u}ez\\
Mayag\"{u}ez, PR, 00680}

\author{Luis Angelats-Silva}
\author{Maharaj Tomar}
\email{maharajs.tomar@upr.edu}
\affiliation{Physics Department\\
University of Puerto Rico at Mayag\"{u}ez\\
Mayag\"{u}ez, PR, 00680}

\date{\today}

\begin{abstract}

The transverse magnetoresistance of thin films of the Diluted Magnetic Semiconductor Zn$_{1-x}$Co$_{x}$O:Al on glass was studied for temperatures in the range of 5 to 100~K. Measurements were made on thin films grown by rf magnetron sputtering, with a thickness of approximately 200~nm. ZnO was alloyed with Co to a concentration $x$ of 0.1 and co-doped with a 5.5\% wt concentration of Al. The electrical resistivity was measured along the sample surface by the four-point probe method with a magnetic field of up to 4~T  applied perpendicular to the surface of the film. The experimental results of the magnetoresistance have been interpreted by means of a semiclassical model that combines a relaxation-time approximation to describe scattering processes in ZnO and a phenomenological approach to the spin-disorder scattering due to the indirect exchange interaction of the magnetic impurities. 
\end{abstract}


\maketitle

\section{\label{sec:Intro}Introduction}

Diluted Magnetic Semiconductor (DMS) materials like Zn$_{1-x}$Co$_{x}$O can be synthesized with a wide variety of transition-metal magnetic impurities, in a wide range of concentrations up to about 35\%. Some ZnO-based DMS have exhibited room-temperature ferromagnetism\cite{Ueda2001, Park2003} as predicted by Dietl and Sato\cite{Dietl2000, Sato2000}, and are of interest for their potential applications in spintronics. 
In doped wide-gap semiconductors charge transport is strongly affected by quantum interference between scattered waves and the wave function amplitudes of the electron-electron interactions. This makes these systems sensitive to phase-breaking mechanisms such as spin relaxation and spin degeneracy,  which can be induced by applying a magnetic field\cite{Andrearczyk2005}. Although there is considerable experimental work on  magnetotransport behavior, like magnetoresistance (MR) and the anomalous Hall effect (AHE) of doped ZnO-based DMS\cite{Dietl2000, Sato2000}, there is no clear consensus yet, due in part to the fact these properties depend, in a complicated manner, on temperature and electron concentration\cite{Xu2007}. The complex superposition of negative and positive MR is not completely understood. In this paper, we report the transverse magnetoresistance of Zn$_{1-x}$Co$_{x}$O:Al thin films on glass. The results on MR have been modeled using a semi-classical model considering relaxation processes and 
sp-d exchange interaction.

\section{Experimental Procedure}
Films were synthesized by RF magnetron sputtering over flint glass substrates with various compositions. Reference samples of ZnO and ZnO doped with Al (ZnO:Al) were made along the with the Zn$_{0.9}$Co$_{0.1}$O:Al 5.5\% wt. All samples were grown at $300\, ^{\circ}$C using an RF power of 125~W under a pressure of $8.5 \times 10^{-3}$ Torr. X-ray analysis showed that the Zn$_{0.9}$Co$_{0.1}$O:Al sample grew along the (002) direction (i.e.\ perpendicular to the c-axis) with a sharp peak at $2\theta = 34.483 ^{\circ}$. The sample thickness of 198~nm was determined by AFM profilometry. 

Resistivity as a function of temperature and applied magnetic field was measured in a four-point-probe arrangement with a Keithley 236 Source/Measure Electrometer with currents down to 1~nA. Gold wire was attached to the film with small drops of silver paint. All three samples showed a good ohmic behavior. At room temperature ZnO, ZnO:Al, and Zn$_{0.9}$Co$_{0.1}$O:Al had resistances of $1.89 \times 10^{8}$, $6.88 \times 10^{5}$, and $1.55 \times 10^{5}\, \Omega$, respectively. The three-order-of-magnitude drop in resistance achieved by Al-doping allowed us to measure resistivity down to 5~K without exceeding the limits of our measurement system.

The temperature and magnetic field were controlled with a Janis 4TL-VT25-4KCCR cryostat equipped with a 4-T superconducting magnet. The resistivity as a function of applied magnetic field was measured at discrete temperatures in the range of 5 to 100~K. The sample was placed in the center of the superconducting solenoid with its surface perpendicular to the direction of the applied magnetic field. 

\section{Semiclassical Model for Magnetoresistance}
We will use the following definition of magnetoresistance:
\begin{equation}
\rho_M = \frac{\rho(H,T)-\rho(T)}{\rho(T)},
\end{equation}
where $\rho(H,T)$ is the resistivity at temperature $T$, under the influence of a magnetic field $H$, and $\rho(T)$ is the resistivity at the same temperature in the absence of a magnetic field. According to Boltzmann's theory, the magnetic-field-dependent resistivity can be expressed as
\begin{equation}
\rho(H,T)/\rho(T) = 1 + \beta_{\bot}H^2 + \beta_{\parallel}\frac{H\left(\vec{j}\cdotp\vec{H}\right)}{\vec{\left|j\right|}},
\end{equation} 
where $\beta_{\parallel}$ and $\beta_{\bot}$ are the parallel and transverse magnetoresistance coefficients, respectively, and $\vec{j}$ is the current density vector. The last term vanishes when the magnetic field is transverse to the current, thus
\begin{equation}
\rho_M = \beta_{\bot}H^2.
\end{equation}
We can now invoke the relaxation-time-approximation to calculate $\beta_{\bot}$
\begin{equation}
\beta_{\bot} = \left(\frac{e}{m_e^{\ast}c}\right)^2\frac{1}{\langle\tau\rangle}\left[\bigg\langle\!{\frac{\tau^3}{1+\gamma^2}}\!\bigg\rangle - \frac{\big\langle{\frac{\tau^2}{1+\gamma^2}}\big\rangle^2}{\big\langle{\frac{\tau}{1+\gamma^2}}\big\rangle}\right],
\end{equation}
where $e$ is charge of the electron, $m_e^\ast$ is the electron effective mass, $c$ is the speed of light, $\langle\tau\rangle$ is the average value of the  relaxation time $\tau$, and $\gamma = \omega_c\tau$. $\omega_c$ is the cyclotron frequency for conduction-band electrons\cite{Blatt1968}.  

Averages are computed semiclassically according to:
\begin{subequations}
\begin{equation}
\langle\mathcal{O}\rangle = \frac{1}{N}{\int_{0}^{\infty}\!\!\!\!\mathcal{O}\left(\!-\frac{\partial f_0}{\partial E}\!\right) E^{3/2}dE},
\end{equation}
\begin{equation}
N = {\int_0^{\infty}\!\!\left(\!-\frac{\partial f_0}{\partial E}\!\right) E^{3/2}dE},
\end{equation}
\end{subequations}
where $f_0$ is the equilibrium Fermi-Dirac distribution. 

We will consider scattering by acoustic phonons, ionized impurities, dislocations, and spin disorder\cite{Blatt1968, Look1989, Look2005}. The latter mechanism will be treated phenomenologically. Only the dominant scattering mechanisms will be used in the calculation. 

Acoustic phonon scattering has a relaxation time
\begin{subequations}
\begin{equation}
\tau_{ph}(E) = a\frac{E^{-\frac{1}{2}}}{k_BT}, 
\end{equation} 
\begin{equation}
a = \frac{2\pi h^4\rho v^2_l}{(2m_e^{\ast})^{\frac{3}{2}}\varepsilon_d^2}, 
\end{equation}
\end{subequations}
where $h$ is Planck's constant, $\rho$ is the mass density, $v_l$ is the speed of sound, and $\varepsilon_d$ is the deformation potential. For ZnO, $a/k_B \approx 57$.

The scattering relaxation time due to dislocations is computed from
\begin{subequations}
\begin{equation}
\tau_{dis}(E) = \frac{\hbar^3\varepsilon c^2}{N_{dis}m_e^{\ast}e^4\lambda^4}\left(1+\frac{8m_e^{\ast}\lambda^2E}{\hbar^2} \right)^\frac{3}{2},
\end{equation}
\begin{equation}
\lambda = \left(\frac{\varepsilon k_BT}{ne^2}\right)^\frac{1}{2},
\end{equation}
\end{subequations}
where $N_{dis}$ is the surface density of dislocations, $\varepsilon$ is the low-frequency dielectric constant, and $n$ is the conduction electron density. 

The ionized impurity scattering contribution is calculated according to
\begin{subequations}
\begin{equation}
\tau_{ii}(E) = \frac{2^{\frac{9}{2}}\pi\varepsilon^2\sqrt{m_e^{\ast}}}{e^4N_I\left[ \ln(1+C_{0}E) - \frac{C_{0}E}{1+C_{0}E} \right]}E^{\frac{3}{2}},
\end{equation}
\begin{equation}
C_{0} = \frac{8\varepsilon m_e^{\ast}k_{B}T}{\hbar^2 e^2 n}, 
\end{equation}
\end{subequations}
where $N_I$ is the total impurity concentration. 

Using values from Look\cite{Look2005} and Madelung\cite{Madelung2004} at $T=100$~K, the relaxation-time averages for the three processes can be estimated: $\langle\tau_{ph}\rangle = 4.68 \times 10^{-5}\: \rm{s},\; \langle\tau_{dis}\rangle = 2.80 \times 10^{-9}\: \rm{s},\; \langle\tau_{ii}\rangle = 1.93 \times 10^{-14}\: \rm{s}$. At lower temperatures $\tau_{ph}$ and $\tau_{dis}$ increase faster that $\tau_{ii}$. This indicates that the limiting nonmagnetic scattering mechanism for ZnO is due to ionized impurities. In our calculations we will ignore both acoustic-phonon and dislocation scattering mechanisms. Spin-disorder scattering will be included as a modification to the ionized impurity scattering. 

\section{Exchange Interaction}

In wide-gap DMS, the indirect exchange interaction between magnetic impurities and band carriers modifies the g-factor such that

\begin{equation}
g_{\alpha, \beta} = g_{\rm ZnO} + x\frac{N_0J\binom{\alpha}{\beta}}{\mu_BH}S\Brill{\!\!\left(\frac{{\!\displaystyle g}_{\rm Co}S\mu_BH}{k_BT}\right)},
\end{equation}
where $g_{\rm ZnO}$ is the g-factor of ZnO, $x$ is the Co concentration, $N_0J\binom{\alpha}{\beta}$ is the exchange integral ($J(\alpha),\, J(\beta)$ for electrons and holes, respectively), $\mu_B$ is the Bohr magneton, $H$ is the applied magnetic field,  $\Brill()$ is the Brillouin function, and  $S$ and $g_{\rm Co}$ are the spin quantum number and Land\'e g-factor for Co$^{+2}$ ions, respectively\cite{Furdyna1988}.

The spin-split Landau levels for the conduction and valence bands relative to the band edges are given by
\begin{subequations}
\begin{equation}
E_c = (n+\frac{1}{2})\hbar\omega_c \pm \frac{1}{2}g_{\alpha}\mu_BH,
\end{equation}
\begin{equation}
E_v = -(n+\frac{1}{2})\hbar\omega_v \mp \frac{1}{2}g_{\beta}\mu_BH,
\end{equation}
\end{subequations}
where $n$ is the index for the Landau level and $\omega_{c}$ and $\omega_{v}$ are the cyclotron frequencies for the conduction and valence bands, respectively. Using the temperature-dependent energy gap for ZnO $E_g(T)$ we can get the temperature- and magnetic-field-dependent gap as
\begin{equation}
E_g(T,H) = E_g(T) + \frac{1}{2}\hbar(\omega_c + \omega_v) - \frac{1}{2}(g_{\alpha} + g_{\beta} )\mu_BH.
\end{equation}

The number of ionized impurities can be expressed as follows, 
\begin{equation}
N_{I} = \left( N_{a}e^{-\frac{E_{a}}{k_BT}} + N_{d}e^{-\frac{E_{d}}{k_BT}}  \right), 
\end{equation} where
$N_{a}$ and $N_{d}$ are acceptor and donor concentrations, respectively and $E_{a}$, and $E_{d}$ are the acceptor and donor binding energies, respectively. We use these as the fitting parameters in the model to calculate $N_{I}$.

A negative contribution to the magnetoresistance can be included phenomenologically as a multiplicative factor for the relaxation time
\begin{equation}
\tau_{\rm eff} = \tau_{ii}\times \left[ \frac{E_g(T,H)}{E_g(T)} \right]. 
\end{equation}
The factor multiplying $\tau_{ii}$ serves to increase the effective relaxation time $\tau_{\rm eff}$ at high magnetic fields. Such behavior is expected from spin-disorder scattering. Its associated relaxation time $\tau_{sd}$, is inversely related to the temperature and the susceptibility\cite{Haas1968, Andrearczyk2005}. As the magnetic field increases, the magnetization saturates as the electrons increasingly populate the lowest spin-split Landau sub-band. This saturation leads to a vanishing susceptibility.

\section{Results and Discussion}
\begin{table*}
\caption{\label{tab:tableI}Parameters of the phenomenological model.}
\begin{ruledtabular}
\begin{tabular}{rrlrr}
$T({\rm K})$ & $n(\times 10^{18}\,{\rm m}^{-3})$ & $N_{I}({\rm m}^{-3})$ & $\tau(\times 10^{-12}\,{\rm s})\footnotemark[1]$ & $\gamma\footnotemark[1]$\\
\hline\\[-6pt]
5 & 5.11 & 1.07$\times$10$^{-23}$ & 78.6000 & 201.260 \\
6 & 5.30 & 6.61$\times$10$^{-19}$ & 5.7000 & 14.600  \\
7 & 6.86 & 1.42$\times$10$^{-11}$ & 2.2600 & 5.790   \\
10& 12.30 & 1.35$\times$10$^{2}$ & 0.0600 & 0.150   \\
15& 26.90 & 1.02$\times$10$^{8}$ &0.0059& 0.015  \\
20& 46.30 & 8.51$\times$10$^{6}$ &0.0587& 0.150   \\
25& 67.40 & 2.05$\times$10$^{11}$ &0.0791& 0.200   \\
50& 206.00  & 3.89$\times$10$^{17}$ &0.0388& 0.099  \\
100&435.00  & 8.94$\times$10$^{20}$ &0.0256& 0.066  \\
\end{tabular}
\end{ruledtabular}
\footnotetext[1]{Calculated at 4~T where $\omega_c = 2.56\times 10^{12}$ s$^{-1}$}
\end{table*}

\begin{figure}
\includegraphics{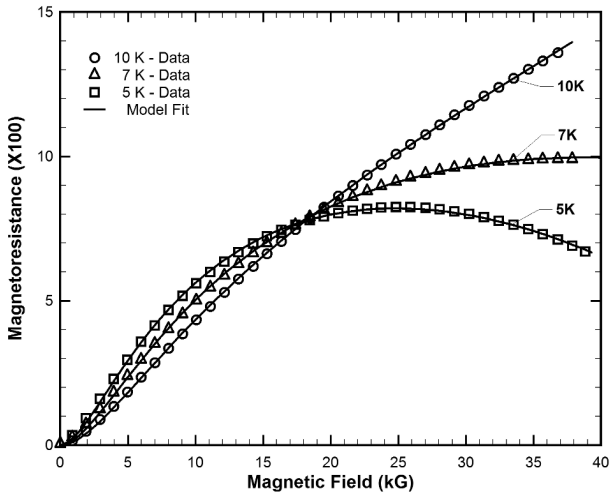}
\caption{\label{fig:5-10K} Magnetoresistance ($\Delta\rho / \rho_o$) at 5~K ($\square $), 7~K ($\triangle $), and 10~K ($\bigcirc$). The model-fit curves are shown as solid lines.}
\end{figure}

Figure~\ref{fig:5-10K} shows the magnetoresistance for temperatures between 5 and 10~K. At 5 and 7~K the positive magnetoresistance dominates at small fields where spin-disorder scattering is still large. When the field has increased enough that the carrier density in the lowest subband is reaching its maximum, the ionized-impurity scattering contribution becomes field independent. The spin-disorder scattering contribution continues to decrease as the applied field increases producing a maximum in the magnetoresistance curve. At 10~K the decrease of the spin-disorder scattering is slower than before since $\tau_{sd} \propto T^{-\frac{3}{2}}\,$\cite{Haas1968}. In this case, the maximum would occur at approximately 41~kG according to an extrapolation of the model.

\begin{figure}
\includegraphics{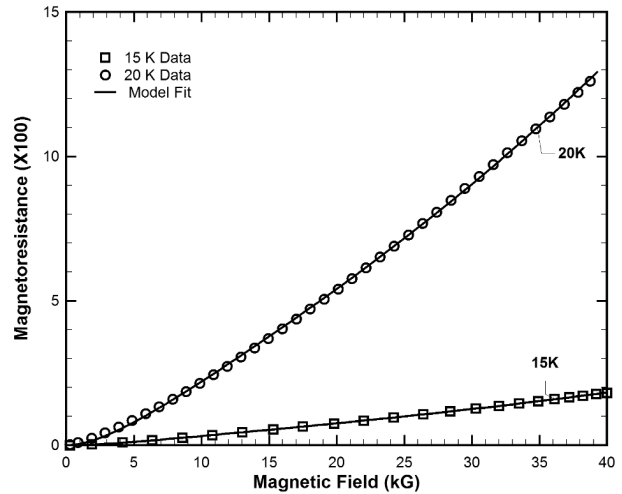}
\caption{\label{fig:15-20K}Magnetoresistance ($\Delta\rho / \rho_o$) at 15~K ($\square $)and 20~K ($\bigcirc$). The model-fit curves are shown as solid lines.}
\end{figure}

At higher temperatures between 15 and 20~K the magnetoresistance increases monotonically as a function of applied field (see Fig.~\ref{fig:15-20K}). At these temperatures, disorder scattering does not decrease enough to compensate for the increase in ionized impurity scattering. There is a sudden drop in the magnitude of the magnetoresistance around 15~K which the fit compensates for but leads to an anomalous increase in $N_{I}$ which cannot be physically accounted for in the model. This result is repeatable and does not appear to be a systematic error. 
Previous studies of magnetoresistance in Zn$_{1-x}$Co$_{x}$O have not shown this behavior\cite{Prellier2003}.

\begin{figure}
\includegraphics{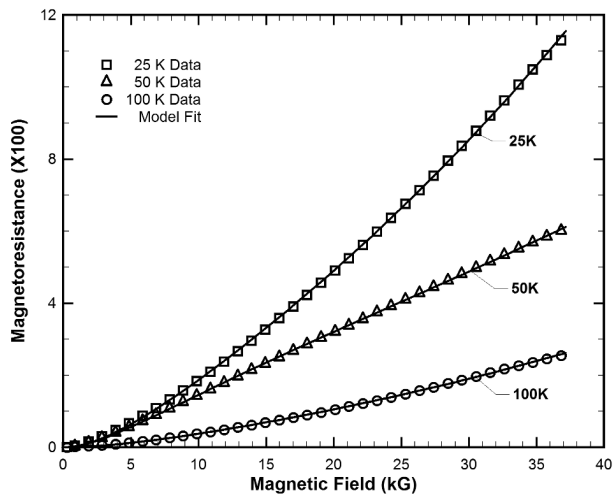}
\caption{\label{fig:25-100K}Magnetoresistance ($\Delta\rho / \rho_o$) at 25~K ($\square$), 50~K ($\triangle$), and 100~K ($\bigcirc$). The model-fit curves are shown as solid lines.}
\end{figure}

As shown in Fig.~\ref{fig:25-100K} between 25 and 100~K the magnetoresistance increases monotonically with $H$ and decreases with temperature. As the temperature increases the number of ionized impurities increases, the band gap decreases and the spin splitting of the Landau subband becomes smaller. As the magnetic field is increased the effective band gap becomes larger again giving rise to a positive magnetoresistance. In this range of temperatures previous authors have observed negative magnetoresistance\cite{Prellier2003}. This might be indicative of a strong s-d exchange interaction between magnetic impurities and band electrons.

\begin{figure}
\includegraphics{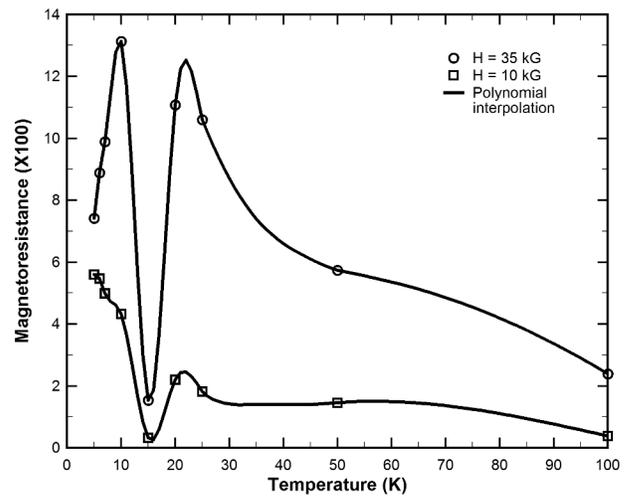}
\caption{\label{fig:10-35kG}Magnetoresistance ($\Delta\rho / \rho_o$) as a function of temperature at 10~kG ($\square $) and 35~kG ($\bigcirc$). The solid lines are polynomial interpolations.}
\end{figure}

Figure~\ref{fig:10-35kG} highlights the overall behavior of the magnetoresistance as function of temperature for two different field values. The salient features are the sharp increase at low temperatures and high magnetic field, the slow decrease at high temperatures, and the anomalous dip at 15~K. This latter feature is unaccounted for by the model used to fit the data in the sense that it produces a fluctuation in the concentration of ionized impurities. 

The parameters obtained from the fit to the phenomenological model are shown in Table~\ref{tab:tableI}.

\begin{acknowledgments}
We wish to acknowledge Dr. Surinder Singh, of UPRM, for his encouragement, valuable suggestions, and enlightened discussions.
\end{acknowledgments}

\bibliography{MRZnO}

\end{document}